\documentclass[prl,aps,twocolumn,superscriptaddress,notitlepage]{revtex4-1}
\usepackage{latexsym,amssymb,amsfonts,amsmath,amstext,graphicx,bbm,bm,hyperref,relsize,dsfont,theorem,times,bigints,bm}
\hypersetup{
	colorlinks=true,
	linkcolor=blue,
	filecolor=magenta,      
	urlcolor=cyan,
}
\usepackage[utf8]{inputenc}

\newcommand{\tr}{\mbox{tr}}
\usepackage{amssymb} 
\usepackage{amsmath} 
\usepackage[utf8]{inputenc} 
\usepackage{array}
\usepackage{makecell}

\def\tr{\mbox{tr}}
\def\bra#1{\langle{#1}|}
\def\ket#1{|{#1}\rangle}

{\catcode`\|=\active
  \gdef\Braket#1{\begingroup
\mathcode`\|32768\let|\BraVert\left<{#1}\right>\endgroup}}
\def\BraVert{\egroup\,\mid\,\bgroup}

\begin{document}

\title{Preparations and Weak Quantum Control can Witness non-Markovianity}
\author{Shlok Nahar}
\affiliation{Department of Metallurgy and Material Sciences, Indian Institute of Technology Bombay, Mumbai 400076, India}
\author{Sai Vinjanampathy}
\email{sai@phy.iitb.ac.in}
\affiliation{Department of Physics, Indian Institute of Technology Bombay, Mumbai 400076, India}
\affiliation{Centre for Quantum Technologies, National University of Singapore, 3 Science Drive 2, Singapore 117543.}

\date{\today}

\begin{abstract}
The dynamics of a system that is initially correlated with an environment is almost always non-Markovian. Hence it is important to characterise such initial correlations experimentally and witness them in physically realistic settings. One such setting is weak-field phase control, where chemical reactions are sought to be controlled by the phase of shaped weak laser pulses. In this manuscript, we show how weak quantum controllability can be combined with quantum preparations to witness initial correlations between the system and the environment. Furthermore we show how weak field phase control can be applied to witness when the quantum regression formula does not apply.
\end{abstract}

\maketitle
\emph{Introduction.---} Almost all open quantum system evolution is non-Markovian. We understand the evolution of a quantum system as non-Markovian when the quantum system exhibits memory of past dynamics. Since we can now control and manipulate small quantum systems coupled to small baths, the evolution of the system is ill-described by models of Markovian dynamics such as Markovian master equations. Examples of such practical non-Markovianity (NM) can be seen in systems with structured environments\cite{gonzalez2017markovian}, quantum biology\cite{ishizaki2010quantum} and NMR \cite{bernardes2016high}, making it ever more relevant to be able to detect NM in an arbitrary physical system.

To proceed with this discussion of NM, we must first define what we mean by it. Several formal definitions of non-Markovianity have been proposed with conceptual similarities and also key differences. Definitions of NM have been formulated based on the non-divisibility of maps \cite{wolf2008dividing,wolf2008assessing,rivas2010entanglement,chruscinski2011measures} and the backflow of information from the environment \cite{Breuer2009}. These measures are part of a hierarchy of NM \cite{Li2017} and are related to the presence of correlations at intermediate times \cite{Fanchini2013}. Another related definition of NM involves the presence of initial correlations \cite{rodriguez2010linear,rodriguez2011non,modi2011preparation,rodriguez2012unification,vinjanampathy2015entropy,vinjanampathy2016correlations}. Such initial correlations violate the assumption of initially factorised states and hence lead to the breakdown of several well known approaches to the dynamics of the reduced state such as Lindblad dynamics and completely positive trace preserving (CPTP) maps \cite{breuer2002theory}. Hence it is imperative that we understand how to characterise and control NM in physical systems for future applications \cite{Mukherjee2015}. One way to characterise NM is to perform state tomography on the system-environment state. Since tomography of the full system-environmental state is impractical and the tomography of non-Markovian systems is cumbersome \cite{pollock2018non} it is important to find methods to witness NM in open quantum systems with minimal assumptions about the system.  

In this manuscript, we demonstrate how weak-field phase control (WFPC) can be used to witness both initial and intermediate correlations. WFPC is a spectroscopic technique where the phase of weak, shaped laser pulses is used to control the dynamics of quantum systems. Examples of weak field phase control of open quantum systems include interesting quantum biological systems such as bacteriorhodopsin \cite{prokhorenko2006} and is the relevant regime of control for the dynamics of protein environments in normal functional conditions. Furthermore WFPC has been shown to be directly influenced by the environment with strong solvent dependence of the stimulated emission \cite{VanderWalle2009}. Computational demonstrations of phase control were presented in \cite{Katz2010,Garcia-Vela2015}. 

We begin by discussing the necessary conditions for observing WFPC before relating it to the correlations between the system and the environment. We then present a method to witness NM induced due to initial or intermediate correlations. Finally, we comment on how the witnessing of correlations can be used as a method to detect physical systems that violate the quantum regression formula. Our results hence connect an important spectroscopic tool that is of importance to quantum physics to the problem of detecting and characterising the nature of NM.

\emph{Model.---}
Consider an open system evolution, where a quantum system $S$ is in contact with an environment $E$. The system is composed of two manifolds, a ground state manifold and an excited state manifold. The control task is to transfer population from the ground state manifold to the excited state manifold. To this end, we define several quantities of interest. First, let $P$ be the projector onto the excited state manifold. Furthermore, let the initial joint state of the system-environment be $R(0)$. The bare Hamiltonian that governs the subsequent evolution is given by $\displaystyle H_{0}$ and a control field that is applied only on the system is given by $V(t)\otimes \mathbb{I}$. Without loss of generality we take the form of the control Hamiltonian ($\hbar=1$) to be composed entirely of off-diagonal blocks in the energy eigenbasis namely
\begin{align}
V(t)=\left( \begin{array}{cc}
0 	& 	\mu \varepsilon (t) \\
\mu \varepsilon^* (t)	 &	 0
\end{array}\right).
\end{align}
Here $\mu$ is a Hermitian operator (such as the dipole moment operator) and $\varepsilon(t)=\mathcal{F}(\tilde{\varepsilon}(\omega))$ is a time dependant field that mediates the external control via the amplitude and phase of $\tilde{\varepsilon}(\omega)=\tilde{A}(\omega)e^{i\varphi(\omega)}$. By phase control, we mean that the expectation value of $P$ is controllable by $\varphi(\omega)$. The state of the system-environment at a later time is given  by a unitary rotation, namely
\begin{align}\label{main}
R(t)=U(t)R(0)U^{\dag}(t).
\end{align}
Here $U(t)=\mathcal{T}\exp(-i\int_{0}^{t}du~\{H_{0}+V(u)\otimes \mathbb{I}\})$. If the initial system-environment state is factorizable as $R(0)=\rho^{(S)}(0)\otimes\tau^{(E)}(0)$, the subsequent evolution of the marginal states is described as a completely positive trace preserving map acting on the initial marginal state alone. We will return to this point later.

The central quantity of interest is the rate of population change in the excited state manifold at time $t$. Consequently, we can define the population rate in the excited state manifold as 
\begin{align}\label{pdot}
\dot{p}(t):=\tr(P\otimes \mathbb{I} \dot{R}(t))
\end{align}
If the field is sufficiently weak, the dynamics of the system is well understood by second order perturbation theory. Following standard literature \cite{kato2013perturbation}, we can write the evolution equation for the joint state at time $t$ in this perturbative regime as 
\begin{align}\label{rdot}
\dot{R}_I(t)=-i [ V_I(t), R_I(0) ]-\int_0^{t}du [V_I(t), [V_I(u), R_I(0)]].
\end{align}
Here, $R_I$ refers to the system environment state in the interaction picture given by $U_0RU^\dag_0=R_I$, with $U_0=exp(iH_0 t)$. The calculation of $\dot{p}(t)$ follows by substituting Eq. (\ref{rdot}) into Eq. (\ref{pdot}) in the appropriate picture. 

We focus on WFPC with an eye to inspect the relationship of the projector $P$ and the bare Hamiltonian, and their relation to the initial system environment state. We begin with a brief summary of the results relating to WFPC before we prove a theorem relating phase control to NM. Hence, in the next section, we will discuss the different conditions for the presence of weak phase control.

\emph{No-Go Theorem for WFPC.---}We begin with the point of view that we have just observed phase control in a physical system. We want to find out what exactly caused this phase control. Am-Shallem and Kosloff \cite{Am-Shallem2014} produced a no-go theorem that asserted that no weak phase control is observed if a set of conditions are met. Since phase control has been observed, one or more of these conditions must have been violated. A generalised version of these conditions are: (1) The field is weak enough for second order perturbation theory to be a good approximation,(2) The free evolution does not excite the system, i.e. $[P\otimes \mathbb{I},H_0]=0$,(3) The initial state is invariant under free evolution, i.e. $[H_0,R(0)]=0$, and (4)Stationarity of the bare evolution, defined as the evolution of the system state $\rho(t_1)$ to $\rho(t_2)$ only depending on the difference $t_2-t_1$.

Condition 1 ensures that the physics excites only the low lying energy sectors of the system, a condition important for several experimental scenarios \cite{prokhorenko2006} where the molecule under consideration can photo-disassociate under strong fields. Condition 2 simply is the statement that the bare evolution should not excite the quantum system to make transitions from the ground state manifold {$\ket{g_i}$} into the excited state manifold {$\ket{e_i}$}. Condition 3 asserts that the initial state should commute with the bare Hamiltonian. This condition has to be understood in the context of the fact that though the off-diagonal terms in the system energy eigenbasis of the type $\ket{g_n}\bra{e_m}$ are important for phase control, not all off-diagonal terms in the energy eigenbasis produce phase control. Consider the bare Hamiltonian $H_0=\omega^{(S)}(n^{(S)}+1/2)+\sum_k\omega^{(E)}_{k}(n^{(E)}_{k}+1/2)+gn^{(S)}\sum_{k}x^{(E)}_{k}$, where the excited state is defined by the projection operator $P=\mathbb{I}-\ket{0}\bra{0}$. Here $\omega^{(S)}$ is the frequency of the system and $\omega^{(E)}_{k}$ represent the frequencies of the environment made of harmonic oscillator modes. If the initial state of the system environment is a Gibbs state given by $R(0)=\exp(-\beta H_0)/Z_0$, where $Z_0=\tr\exp(-\beta H_0)$, then all conditions are satisfied and there is indeed no phase control. Note that the absence of phase control survives the initial correlations between the system and environment, but only because the off diagonal terms (in the energy eigenbasis of the individual system and environment spaces) of  $R(0)$ are all in the environment.

Condition 4 deals with stationarity, which is not important for our discussion. The dynamics of quantum systems whose system-environmental states are factorizable can always be made stationary \cite{Chruscinski2010}. On the other hand, when a quantum system possesses initial correlations, the underlying assumption being made is that the bath correlation functions do not decay sufficiently quickly. In this case, the dynamics of the system and environment are inseparable and stationarity does not hold. In such cases though, the discussion (say Eq.(12) and Eq.(13) in \cite{Am-Shallem2014})  can be restated such that the results are unaffected.

Having discussed Conditions 1 and 2 above, we inspect the other conditions  of the no-go theorem to see when the detection of WFPC can witness the non-Markovian evolution of the quantum system. Given that the other condition is a statement about the initial state and its evolution, we inspect the initial state of the system and the map that evolves this system to the final state. 

In our notation, the projector is given by $P=\sum_{i}\ket{e_i}\bra{e_i}$ and the control Hamiltonian can be written as 
\begin{align}
V(t) = \sum_{jk} \mu_{kj}\varepsilon(t)\ket{e_j}\bra{g_k} + \mu_{jk}\varepsilon^*(t)\ket{g_k}\bra{e_j}
\end{align} 
The general system-environment state can be written in terms of the ground and excited system state manifolds as
\begin{widetext}
\begin{align} \label{r0}
R(0) =  \sum_{i,j,k,l} a^{(0)}_{jklm}\ket{g_m}\bra{g_k}\otimes\ket{\alpha_j}\bra{\alpha_l} +  b^{(0)}_{jklm}\ket{e_m}\bra{e_k}\otimes\ket{\alpha_j}\bra{\alpha_l}
 +c^{(0)}_{jklm}\ket{g_j}\bra{e_k}\otimes\ket{\alpha_l}\bra{\alpha_m}+H.c.
\end{align}
\end{widetext}
where $ \{a^{(0)}_{jklm}, b^{(0)}_{jklm}, c^{(0)}_{jklm}\}$ are the coefficients of the different operators such that the matrix $R(0)$ is a good density matrix and $\{\ket{\alpha_k}\}$ is a basis for the environmental states. Note that one of the conditions for it to be a good density matrix is that $R(0)$ cannot have any terms of the form $\ket{g_m}\bra{e_k}$ without having both $\ket{g_m}\bra{g_m}$ terms and $\ket{e_k}\bra{e_k}$ terms. Since we want the initial state of the system to be block-diagonal in the energy eigenbasis, we set $a^{(0)}_{jklm}=0$ and write 
\begin{align} \label{Ronnie}
	\nonumber R(0) = & \sum_{jklm} a^{(0)}_{jklm}\ket{g_m}\bra{g_k}\otimes\ket{\alpha_j}\bra{\alpha_l}\\
	& +b^{(0)}_{jklm}\ket{e_m}\bra{e_k}\otimes\ket{\alpha_j}\bra{\alpha_l}
\end{align}
If all the above mentioned conditions are met and the initial state is of the form as in Eq. (\ref{Ronnie}), then phase control is not possible.\\

\emph{Violation of Conditions.---} To summarise the discussion so far, WFPC can be observed in a chemical reaction for several reasons. The first reason is that second order perturbation theory is not valid. This is discounted since control fields are accessible to experimental observation and hence can always be made weak. Furthermore, if we consider applications such as photoisomerization, the important states of interest lie in the lower excited energy subspaces and hence stronger fields would move the system out of a region of interest \cite{prokhorenko2006}. 

The second reason is that free evolution operator does not commute with the target operator. For example, consider the bare Hamiltonian $H_0=\omega^{(S)}(n^{(S)}+1/2)+\sum_k\omega^{(E)}_{k}(n^{(E)}_{k}+1/2)+gx^{(S)}\sum_{k}x^{(E)}_{k}$ where the symbols are defined as before. Furthermore, assume that the system and environment in a  Gibbs state at temperature $\beta^{-1}$, i.e., $R(0)=\exp(-\beta H_0)/Z_0$. Since such a state commutes with $H_0$, it satisfies condition 3. Clearly $[H_0,P]\neq0$, and such an objective (the population in the excited state manifold) is phase controllable following \cite{Am-Shallem2014}. The essence of phase controllability for this example lies in the fact that the system-environment Hamiltonian (which is proportional to the coupling constant $g$) is indeed off-diagonal in the energy eigenbasis of the system. This was pointed out in \cite{Pachon,Spanner2010}, where the authors made the argument that the environment can assist in phase controllability by having a generic system-environment Hamiltonian that does not commute with the target operator. We assume Condition 2 to be true so that we can witness NM using WFPC. The violation of Condition 2 is subsumed into one of the four experimental outcomes discussed below.

Given the condition that the bare evolution does not transfer population from the ground to excited state manifold, the control Hamiltonian in the interaction picture becomes 
\begin{align}
	\nonumber V_I(t)= & U(t)^{\dag}(V(t)\otimes \mathbb{I})U(t)\\
	V_I(t)= & \sum_{jklm}\tilde\mu_{mj}\varepsilon(t)\ket{e_j}\bra{g_m}\otimes\ket{\alpha_k}\bra{\alpha_k} + H.c.
\end{align}
Here, $\tilde{\mu}_{jk}$ is the matrix element of the operator $\mu$ in the interaction picture. We can show that the phase control arises from the off-diagonal blocks in the energy eigen basis. To this end we consider the initial state to be as in Eq. (\ref{r0}) 
Taking the first order term from Eq. (\ref{rdot}) and putting it in Eq. (\ref{pdot}), we obtain $\dot{p} = -it_\alpha \sum_{lmn}( \tilde\mu_{ml}c_{lmnn}\varepsilon(t)-\tilde\mu^*_{ml}c^*_{lmnn}\varepsilon^*(t))$, which clearly depends on phase.
  Here $t_\alpha$ is the trace over the environmental state. This phase controllability comes entirely from the off-diagonal blocks in the system space of the initial density matrix as shown in Appendix B.
For completeness, we show in Appendix A that if we start with an initial state that is diagonal in the energy eigen basis, there is no phase control. 

\begin{figure}[ht]
	\resizebox{0.8\columnwidth}{!}{%
		\includegraphics{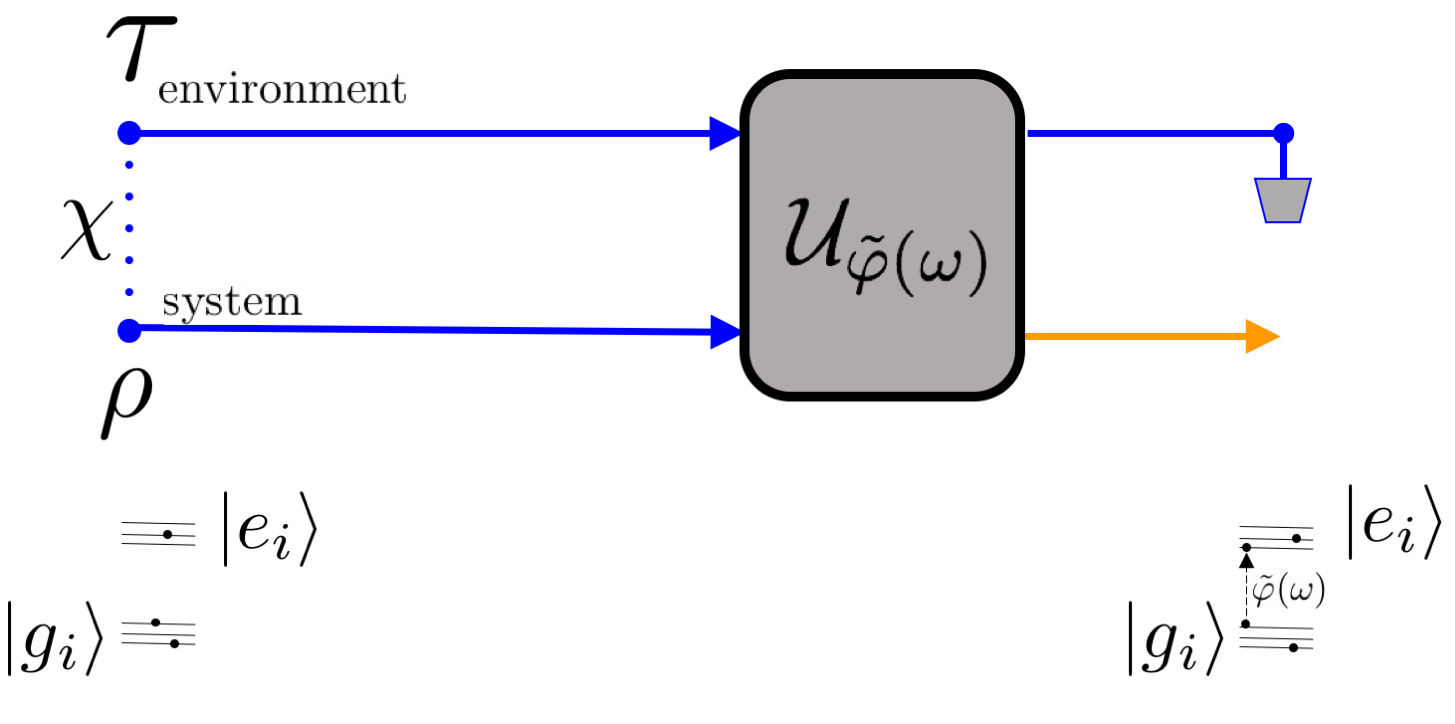}}
	\caption{Weak-Field Phase Control: The system is depicted with a ground state manifold and an excited state manifold. An initially correlated system-environment state $R(0)=\rho\otimes\tau+\chi$ is subject to a joint unitary operator that depends on phase $\varphi(t)=\mathcal{F}[\tilde{\varphi}(\omega)]$ of a control field $\varepsilon(t)$. The system is said to be weak-field phase controllable if the population in the excited state manifold can be controlled by the phase of sufficiently weak control pulses.} 
	\label{fig:algorithm} 
\end{figure}

Condition 3 deals entirely with the initial state of the system and is at the heart of the phase control that shall help us detect the NM implied by the presence of initial correlations. Since we established that phase control arises from off-diagonal terms such as $\ket{g_m}\bra{e_n}$, we seek to place such terms either in the initial system marginal $\rho$  or the initial correlations $\chi$. Distinguishing these two scenarios would directly lead us to the detection of NM. To this end we will consider quantum preparations. 

\emph{Witness of non-Markovianity.---}  If a quantum system is initially correlated with the environment, then operations on the state alone are ill-defined unless the effect of such operations on the environment are accounted for. In this context, we discuss preparations, which are colloquially understood as the act of projecting a system onto a known state usually referred to as the initial state. More precisely, a preparation is a map from an unknown quantum state to a known quantum state \cite{Modi2012,vinjanampathy2015entropy,vinjanampathy2016correlations}. For example, the ``throw and replace" preparation is given by the action $\mathcal{A}_{1}[\rho]=\rho_0$ $\forall \rho\in\mathcal{B}(\mathcal{H})$ and simply maps any initial marginal state of the system to a fixed state $\rho_0$. 
%
%
  Another example of a preparation is the disentanglement channel or the ``marginal preserving" preparation. The action of such a preparation is to disentangle a given joint state, namely $\mathcal{A}[R(0)]=\rho\otimes\tau$ which decorrelates the system and the environment \cite{terno1999nonlinear,DAriano2008}. This preparation cannot be carried out universally with just one density matrix \cite{terno1999nonlinear} but can be performed easily with two copies of the system-environment density matrices as shown in Appendix C.

Phase control can arise from off-diagonal terms of the type $\ket{g_m}\bra{e_n}$ arising either in the marginal state of the system or the correlation matrix, or both.  To detect where phase controllability arises from, we consider marginal preserving preparations $\mathcal{A}$ defined above. We note that before $\mathcal{A}$, the joint system-environment state is given by $R(0)=\rho\otimes\tau+\chi$ whereas after $\mathcal{A}$, the joint system-environment state is given by $R(0)=\rho\otimes\tau$. If all of the off-diagonal terms are present in $\chi$, the marginal preserving preparation $\mathcal{A}$ erases the correlation matrix, thus removing weak-field phase controllability from the system. On the other hand, if the off-diagonals are all present in $\rho$, then the weak field phase controllability is not disrupted by $\mathcal{A}$. Note that though the reaction yield (whose rate is given by $\dot{p}$) changes because $\dot{p}$ depends on $\chi$ in general, the phase dependence of $\dot{p}$ does not change because all of the control is attributed to the marginal state. 

\begin{figure}[ht!]
\resizebox{0.8\columnwidth}{!}{%
\includegraphics{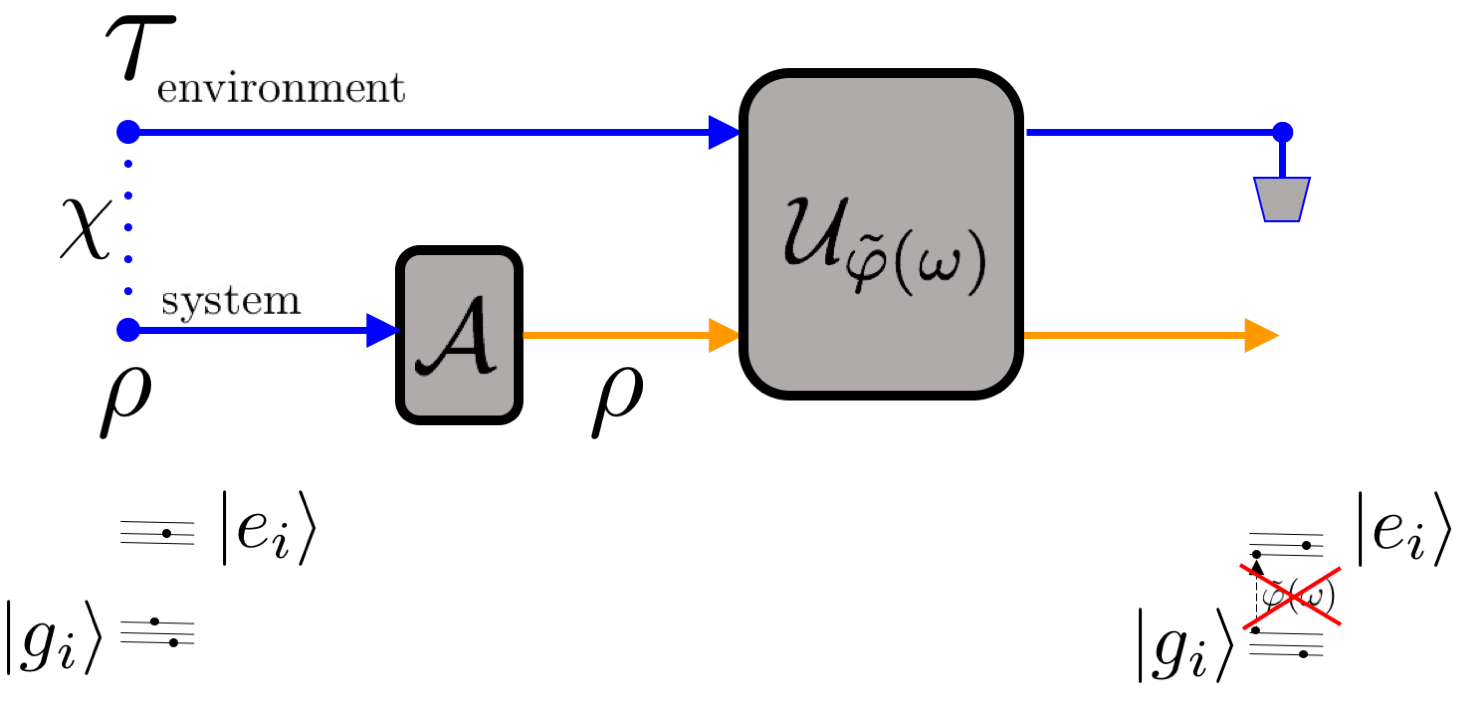}}
\caption{Witness of non-Markovianity: A marginal preserving preparation $\mathcal{A}$ can be used to witness initial correlations between the system and the environment, denoted by $\chi$. The preparation $\mathcal{A}$ witnesses correlations since the output of a marginal preserving preparation is an uncorrelated state, and hence any existing correlations can be seen in the phase controllability of the reaction products.}
\label{fig:witness} 
\end{figure}

We propose an experiment performed on two copies \cite{two_copy} of the system-environment state that can witness NM. The first copy is simply checked to see if the system enjoys WFPC. On the second copy we perform the marginal preserving preparation on the system, and once again check for WFPC. If we detect WFPC before $\mathcal{A}$ and no WFPC after $\mathcal{A}$, then we have detected the correlation matrix whose off-diagonal terms induced WFPC. This witnessing of the correlations is illustrated in Fig. (\ref{fig:witness}). On the other hand, if we detect no WFPC both before and after the marginal preserving preparation $\mathcal{A}$, then we cannot say that there were no correlations between the system and the environment. In Appendix D, we show that the set of all system-environment density matrices which have a non-zero correlation matrix that cannot be detected by the dual WFPC test outlined above is measure zero. 

We also consider the scenario where the joint state $R(0)$ has the relevant off-diagonal terms in both the marginal system state and the correlation matrix. In this case  the reaction yield $p(t)$ is phase controllable, but this phase controllability arises due to both aforementioned terms. This means that the marginal preserving preparation, which removes the correlation matrix $\chi$ will change the quantitative details of the phase controllability. This quantitative change can witness the presence of off-diagonal elements in both terms, as detailed in Appendix B. For completeness we also consider the case when the off-diagonal terms are in the marginal system state but not in the correlation matrix. In this case the quantitative details of the phase controllability will not change after the marginal preserving preparation. We summarise this in table 1.
\begin{table}
  \caption{Table Summarising the Witnessing of non-Markovianity} \label{tab:title} 
\begin{center}
  \begin{tabular}{ | l | c | r |}
    \hline
   Off-Diagonals & Not in $\chi$ &     In $\chi$ \\ \hline
Not in $\rho$ & No WFPC & WFPC$\rightarrow$ No WFPC\\ \hline
In $\rho$ & \makecell{$\frac{dp(t)}{d\varphi}$ unchanged between \\ experiments} &  \makecell{$\frac{dp(t)}{d\varphi}$ changes between \\ experiments}\\
    \hline
  \end{tabular}
\end{center}
\end{table}

Finally, we consider the correlation matrix at intermediate times and apply this formalism to the quantum regression formula (QRF). Consider an initial system-environment state given by $R(0)=\rho(0)\otimes\tau(0)$ evolving to an intermediate time as $R(t_1)=U_0(t_1)R(0)U^{\dag}_0(t_1)$. Here $U_0(t)$ is the ``free" evolution of the system-environment before the laser field $\varepsilon(t)$ is switched on. Now, the two-time correlation function is given by 
$\langle B(t_2)A(t_1)\rangle=\tr( U^{\dag}_0(t_2)BU_0(t_2)U^\dag_0(t_1)AU_0(t_1)R(0) )$ \cite{guarnieri2014quantum}. This expectation value can be written as $\tr_S(B\mathbb{Z})$, where $A, B$ are the system operators in the Schr\"{o}dinger picture and $\mathbb{Z}$ is given by
\begin{align}
\mathbb{Z}=\tr_E(U_0(t_2-t_1)AR(t_1)U^{\dag}_0(t_2-t_1)).
\end{align}
If $R(t_1)=\rho(t_1)\otimes\tau(t_1)$ then $\mathbb{Z}=\Phi_{t_1\rightarrow t_2}[A\rho(t_1)]$, where the CPTP map $\Phi_{t_1\rightarrow t_2}$ is the evolution operator for the system dynamics. Hence the evolution of the two-time correlation function is governed by the same evolution equation as the density matrix, following the spirit of Onsager's regression theorem. If $R(t_1)= \rho(t_1)\otimes\tau(t_1)+\chi(t_1)$, then the regression formula immediately breaks down. Thus QRF relies on the fact that for all $t_1<t_2$, the total state at the intermediate time is assumed to be well approximated by a product state $\rho(t_1)\otimes\tau(t_1)$. The validity of this stronger ``factorization" approximation \cite{Li2017} needs to be ascertained before QRF can be applied to a given physical system. Such violation of the QRF is understood to be yet another definition of NM.

If we partition the physical system under consideration to have a similar structure (a ground and an excited state manifold) and furthermore if the system is not spontaneously excited by the free evolution $U_0(t)$, then following our discussion, WFPC can detect this dynamical definition of NM. Clearly, if we switch on a weak laser field at the intermediate time $t_1$, following our earlier discussion, we can witness the correlation matrix at intermediate times.

\emph{Conclusions.---} Our ability to detect and charecterise NM in physical systems is crucial to understand complex quantum systems. While some aspects of NM arise due to the dynamics of the system and the environment, the presence of initial correlations can also lead to non-Markovian dynamics of the system. In this manuscript, we show how standard spectroscopic techniques can be combined with preparations to witness NM that is induced by the presence of both initial and intermediate correlations. This is important since often the dynamical modelling of a physical system follows assumptions such as the Born-Markov approximation. Such assumptions make strong claims about the nature of the initial and subsequent density matrix of the system-environment, which can be now be checked by preparations and weak quantum control. Furthermore, the applicability of important theorems such as the generalised quantum regression formula rely on the absence of intermediate correlations. The presence of (almost all) such correlations can be witnessed by making a small set of experimentally verifiable assumptions.

In our analysis, we assumed that the free evolution of the system and environment does not excite the system. If this condition is violated, then the witness part of our experiment is unaffected. This is because we are detecting the effect of discarding the correlation matrix on WFPC when we inspect the two parts of the experiment. When this condition is also violated, we cannot differentiate the WFPC that arises from the presence of off-diagonals in the system state alone from the phase control that arises from the violation of the condition. Thus we lose the ability to detect the off-diagonal elements of $\rho$ but this is easily overcome by doing tomography on the system state, which is typically much smaller than the environment.

Finally, we note that WFPC is used here as an alternative to quantum process tomography on initially correlated systems \cite{pollock2017tomographically,milz2016reconstructing}. The full reconstruction of the dynamical map for correlated dynamics typically scales very unfavourably with the size of the universe. Such proposals will herald new experimental progress in the detection, characterisation and control of non-Markovian systems.

The authors thank Ronnie Kosloff for detailed discussions about the manuscript. S.V. acknowledges support from an IITB-IRCC Grant No. 16IRCCSG019 and by the National Research Foundation, Prime Minister's Office, Singapore under its Competitive Research Programme (CRP Award No. NRF-CRP14-2014-02).

\begin{thebibliography}{36}
\bibitem [{two()}]{two_copy}%
  \BibitemOpen
  \href@noop {} {}\bibinfo {note} {Instead of two copies, we can perform the
  experiment sequentially by resetting the entire system environmental
  state.}\BibitemShut {Stop}%
\end{thebibliography}

%

	\section{Appendix A: Proof that Excited states do not produce WFPC}
	Consider an initial state $R(0)=\sum_{jk}\ket{e_j}\bra{e_j}\otimes \ket{\alpha_k}\bra{\alpha_k}$ which has diagonal states in the excited state manifold, without loss of generality. The first order term in the perturbative series expansion will clearly be 0 as 
	\begin{align}
	[V_I(t),R_I(0)]=\sum_{jkl}\tilde\mu^*_{lj}\varepsilon^*(t)\ket{g_l}\bra{e_j}\otimes \ket{\alpha_k}\bra{\alpha_k}-H.c.
	\end{align}
	This has no diagonal terms and hence its trace is zero. To check for phase control in the second order, we evaluate it to be 
	\begin{widetext}
		\begin{align}
		\nonumber& [V_I(t),[V_I(u),R_I(0)]]
		\displaystyle= \sum_{jklm}\tilde\mu^*_{lj}[\tilde\mu_{lm}\varepsilon(t)\varepsilon^*(u)\ket{e_m}\bra{e_j}\otimes\ket{\alpha_k}\bra{\alpha_k}
		\nonumber-\tilde\mu_{mj}\varepsilon(t)\varepsilon^*(u)\ket{g_l}\bra{g_m}\otimes\ket{\alpha_k}\bra{\alpha_k}]
		+ H.c.
		\end{align}
	\end{widetext}
	Acting on this by the projection matrix and performing trace, we get 
	\begin{flalign}
	t_\alpha\sum_{lj}|\tilde\mu_{lj}|^2\varepsilon(t)\varepsilon^*(u)+c.c.
	\end{flalign}
	Where $t_\alpha$ is the trace of the environment marginal state.
	We can summarise the result as 
	\begin{flalign}
	\dot{p} = t_\alpha\int_{0}^{t}du\sum_{lj}|\tilde{\mu}_{lj}|^2\varepsilon(t)\varepsilon^*(u)+|\tilde{\mu}_{lj}|^2\varepsilon^*(t)\varepsilon(u)
	\end{flalign}
	This is clearly dependent on the auto-correlation function and has been proved to be independent of the phase by Am-Shallem and Kosloff \cite{Am-Shallem2014}.	
	\section{Appendix B: Proof of Phase controllability from off-diagonal blocks}
We have already shown in Appendix A that the diagonal terms of the density matrix R(0) as defined in Eq.~(\ref{r0}) do not contribute to phase control. Hence, consider the off-diagonal terms of $R(0)$, namely  $\sum_{lmnk}c_{lmnk}\ket{g_l}\bra{e_m}\otimes \ket{\alpha_n}\bra{\alpha_k}+H.c.$ Here we shall only analyse the off-diagonal terms.
\begin{widetext}	
	\begin{flalign}
		\nonumber [ V_I(t), R_I(0) ]= & \sum_{ilmnk}\tilde\mu_{il}c_{lmnk}\varepsilon (t)\ket{e_i}\bra{e_m}\otimes\ket{\alpha_n}\bra{\alpha_k} - \tilde\mu^*_{im}c_{lmnk}\varepsilon^* (t)\ket{g_l}\bra{g_i}\otimes\ket{\alpha_n}\bra{\alpha_k}- H.c.\\
		\nonumber P\otimes \mathbb{I} [ V_I(t), R_I(0) ]= &  \sum_{ilmnk}\tilde\mu_{il}c_{lmnk}\varepsilon (t)\ket{e_i}\bra{e_m}\otimes\ket{\alpha_n}\bra{\alpha_k}- H.c.\\
		tr(P\otimes \mathbb{I}\dot{R}_I(t))= & -itr( P\otimes \mathbb{I}[ V_I(t), R_I(0) ] )\\
		\dot{p} = & -it_\alpha \sum_{lmn}( \tilde\mu_{ml}c_{lmnn}\varepsilon(t)-\tilde\mu^*_{ml}c^*_{lmnn}\varepsilon^*(t))
	\end{flalign}
\end{widetext}		
	This clearly depends on the phase.
	Additionally, let us consider the case when some of these off-diagonal terms were in the marginal system state and some were in the correlation matrix. After conducting the two experiments, the terms from the correlation matrix would not be there any more due to the marginal preserving preparation as described in the paper. This would clearly change $\frac{dp(t)}{d\varphi}$ as there would be less terms in the summation.
	
	Another approach to detecting the correlation matrix when both the marginal system state and the correlation matrix have $\ket{g_m}\bra{e_n}$ terms is as follows. We first make two separate preparations of the system into different states $\ket{\psi_m}\bra{\psi_m}\otimes \tau^{E|\psi_m}$ where $\psi_m \in g,e$ and $\tau^{E|\psi_m}$ is the marginal environmental state given that the system is in $\psi_m$. Note that if there are no correlations, the marginal environmental state is the same for all such preparations of the system state. We then rotate this prepared state by a unitary $L$ such that the resulting system state has off-diagonal terms of the form $\ket{g_m}\bra{e_n}$. We can then check the amount of phase control from both our prepared and rotated states. If both of these states have different environmental marginals, then the amount of phase control would be different as the trace over the environment $t_\alpha$ would be different. Thus, this would be a witness of initial correlations.
\section{Appendix C: Marginal Preserving Preparation}
Marginal preserving preparations (MPP) refer to a general decorrelating map that takes a bipartite quantum system as input and outputs the marginal states. As shown in \cite{terno1999nonlinear,DAriano2008}, this MPP cannot be universal for a single copy of the system. In the figure we illustrate the marginal preserving preparation given two copies of the system with the environment. By construction, we hence show that a universal MPP is possible if two copies are available.
	 \begin{figure}[ht!]
		\resizebox{0.5\columnwidth}{!}{%
			\includegraphics{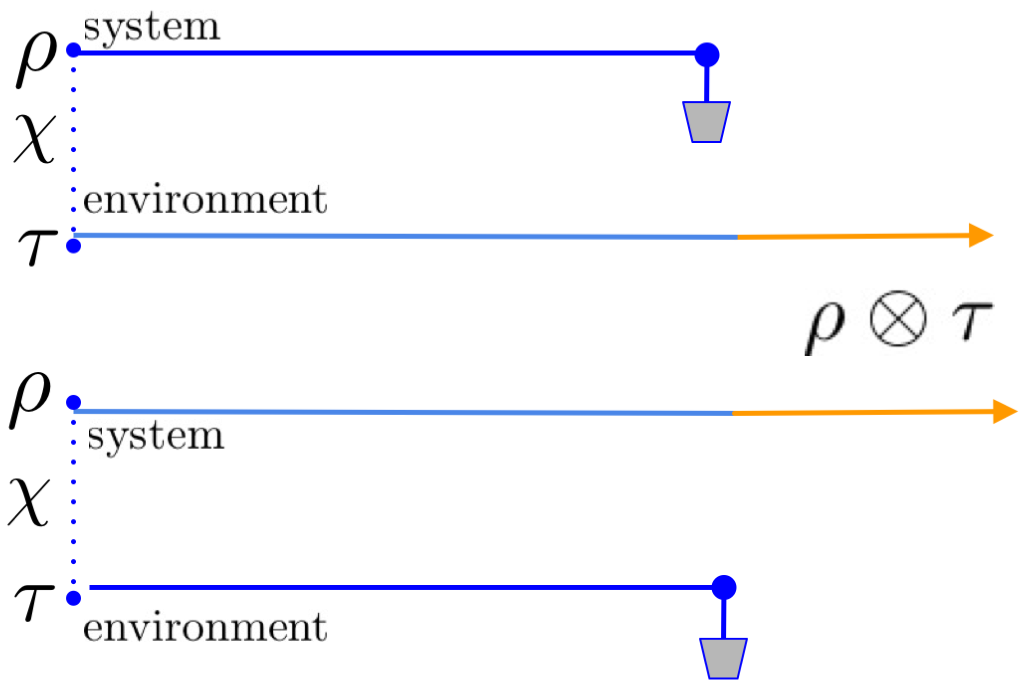}}
		\caption{Marginal Preserving Preparation illustrated with two copies of the system-environment state. By swapping the state of the system from one copy to another, the system state is no longer correlated with its new environment while preserving the marginal states.} 
		\label{fig:algorithm} 
	\end{figure}
	\section{Appendix D: The set of all $\chi$ with no $\ket{g}\bra{e}$ terms is a set of measure zero}
	We express a general correlation matrix in the energy eigen basis of the system as
	\begin{align}
	\nonumber \chi=\sum_{lmnk} a_{klmn}\ket{g_k}\bra{g_l}\otimes\ket{\alpha_m}\bra{\alpha_n}\\
	\nonumber + b_{klmn}\ket{g_k}\bra{e_l}\otimes\ket{\alpha_m}\bra{\alpha_n}\\
	\nonumber + c_{klmn}\ket{e_k}\bra{e_l}\otimes\ket{\alpha_m}\bra{\alpha_n}\\
	+ H.c.
	\end{align}
	For there to be no $\ket{g}\bra{e}$ terms, 
	\begin{align}
	b_{klmn}=0\  \forall \  k,l,m,n
	\end{align}
	With just one of these constraints we have the remaining sub-set of $\chi$ confined to a lower dimensional subspace of all $\chi $. Thus, the set of $\chi$ without any $\ket{g_m}\bra{e_n}$ terms has Lebesgue measure zero.

\end{document}